# A penalized method for multivariate concave least squares with application to productivity analysis [1]

Abolfazl Keshvari

*Aalto University School of Business, Helsinki, Finland*

abolfazl.keshvari@aalto.fi, tel: 00358503120915

**Abstract**

We propose a penalized method for the least squares estimator of a multivariate concave regression function. This estimator is formulated as a quadratic programming (QP) problem with $O(n^2)$ constraints, where $n$ is the number of observations. Computing such an estimator is a very time-consuming task, and the computational burden rises dramatically as the number of observations increases. By introducing a quadratic penalty function, we reformulate the concave least squares estimator as a QP with only non-negativity constraints. This reformulation can be adapted for estimating variants of shape restricted least squares, i.e. the monotonic-concave/convex least squares. The experimental results and an empirical study show that the reformulated problem and its dual are solved significantly faster than the original problem. The Matlab and R codes for implementing the penalized problems are provided in the paper.

## 1. Introduction

This paper is concerned with the shape restricted least squares problem, which is used to estimate a concave or convex regression function. Such an estimator is used in different disciplines: such as productivity analysis (Keshvari & Kuosmanen, 2013; Kuosmanen, 2012; H. Varian, 1984), econometrics (Aït-Sahalia & Duarte, 2003; H. R. Varian, 1982), statistics (Birke & Dette, 2007; Hanson & Pledger, 1976; Hildreth, 1954), and operations research (Badinelli, 1986; Zhou & Lange, 2013).

The estimated function is selected among all the possible functions satisfying the shape assumption. The function is shown to be a piecewise linear function, and it is formulated as a quadratic programming (QP) problem (Kuosmanen, 2008). The finite sample properties of the shape restricted least squares estimator are known. For example, it is known that it satisfies the orthogonality condition, and the mean of fitted values is equal to the mean of the responses. The properties and characteristics are studied by several researchers (see for example Groeneboom, Jongbloed, &

---

Wellner, 2001; Hanson & Pledger, 1976; Kuosmanen, 2008; Mammen, 1991; Meyer, 2003, 2006; Nemirovskii, Polyak, & Tsybakov, 1985; Seijo & Sen, 2011).

Single variate problems are relatively easy to solve and several methods are proposed to compute the estimator (Dykstra & Robertson, 1982; Dykstra, 1983; Hanson & Pledger, 1976; Hildreth, 1954). However, dealing with a multivariate problem is difficult and it is very time consuming. The main source of the computational burden is the number of constraints that is of order $O(n^2)$ and rises very quickly as the number of observations ($n$) increases. To solve this problem, Holloway (1979) proposed an iterative algorithm that approximates the regression function, and it is applied on small size samples. Another approach is proposed by Fraser & Massam (1989) and Meyer (1999) that is a mixed primal–dual algorithm to find a least squares regression estimate over the closed convex cone defined by the constraints. Goldman and Ruud (1993) also propose a generalization to the algorithms of Hildreth (1954) and Dykstra (1983). However, these algorithms are not practical for multi-input problems.

The least squares concave or convex function is piecewise linear consisting of several linear segments. In applications, only a relatively small percentage of the constraints of the related QP are binding and as a result, the number of linear segments is smaller than the number of observations. This result is recently used as the basis for two methods. One of the methods is to preprocess the problem based on the Dantzig's relaxations method (G. B. Dantzig, Fulkerson, & Johnson, 1959; G. Dantzig, Fulkerson, & Johnson, 1954) and to iteratively eliminate some of the nonbinding constraints (Lee, Johnson, Moreno-Centeno, & Kuosmanen, 2013). Based on the pairwise distance between observations, in every iteration a subset of constraints is selected and then a QP is solve to get the solution of the relaxed problem. The other method is to find acceptable partitions of the input space, and to estimate the linear segments for the partitions (Hannah & Dunson, 2013), which may end up with an approximation of the optimal solution. Both of these methods are iterative algorithms and it is required to implement special codes for using them.

In this paper, a reformulation to the QP problem is proposed. In this method, the constraints, except the signs of the variables, are eliminated and the objective function is penalized by the constraints' violations. The final problem is a QP with only sign constraints, and it is solved in a reasonably shorter time than the original problem. To this end, first we convert the original problem into a QP with equality constraints, and categorize the constraints into $n$ blocks of equations. Then the errors are estimated from the first block, and the objective is penalized by the sum of the quadratic values of violations. The dual of the penalized problem is also developed. The dual problem is a separable QP and it is solved significantly faster than the penalized and the original problems. Moreover, a similar approach is used to develop the penalized problem and its dual for estimating variants of shape restricted least squares functions, i.e. the (monotonic) convex and concave least squares.

Since the seminal work of Fiacco and McCormick (1968), the penalty method is well studied in the literature of optimization (e.g. Di Pillo & Grippo, 1989; Hu & Ralph, 2004; Li, Yin, Jiang, & Zhang, 2013). Penalty method is used to solve a wide range of regression problems. For example, Ridge regression (Hoerl & Kennard, 1970) is a penalty method that regularizes coefficients to control their variances. Lasso (Tibshirani, 1996) is a shrinkage and selection method that enhances the out-of-sample interpretability of a regression problem. Moreover, various penalty methods are developed to solve constrained optimization problems, such as in bilevel programming problems (Marcotte & Zhu, 1996), options pricing (D'Halluin, Forsyth, & Labahn, 2004), and portfolio optimization (Corazza, Fasano, & Gusso, 2013). To the best of our knowledge, this paper is the first application of penalty method to solve shape restricted least squares.

To compare the efficiency of the penalized problem and its dual, a number of Monte Carlo simulations is used. The results show the superiority of the penalized shape restricted least squares and the dual problem over the conventional formulation in terms of the computational time. The results show that solving the dual problem is the most efficient approach.

The rest of the paper is organized as follows. Section 2 contains the main results of the paper. It starts with a review on the monotonic concave least squares problem. The steps to build the penalty term, the penalized and the dual problems are explained in this section. The optimality of the penalized problem is also discussed. Moreover, the penalization method for estimating a concave function is developed. To simplify the algebraic calculations, this paper uses succinct matrix forms of the problem. The summation form of the dual problem is also presented in this section. The results of the numerical Monte Carlo simulations are presented in Section 3. An empirical application is presented in Section 4, in which the penalized method is used to analyze the room rates of a sample of hotels in Finland. The paper has three appendices. The first appendix presents the detailed analytical computations to obtain the matrix form of the QP. This appendix also includes the steps to compute the reformulated problem. The second appendix contains the proofs of the theorems. Appendix 3 presents the Matlab and R codes for solving the penalized monotonic concave least squares and the dual problem.

## 2. Penalized monotonic concave least squares

Our focus in this paper is on the concave least squares (CLS) problem under monotonicity assumption (MCLS).[2] One of the applications of this problem is to estimate a non-parametric production function that is monotonic and concave (Andor & Hesse, 2014; Cheng, Bjørndal, & Bjørndal, 2014; Eskelinen & Kuosmanen, 2013; Keshvari & Kuosmanen, 2013; Kuosmanen, 2011, 2012; Wang, Wang, Dang, & Ge, 2014).

### 2.1 Monotonic concave least squares (MCLS)

MCLS is a least squares approach to construct a non-parametric multivariate regression model. In this model, a function $f: \mathbb{R}^m \to \mathbb{R}$ is estimated as $y_i = f(\mathbf{x}_i) + \varepsilon_i$ ($i = 1, ..., n$), where $n$ is the number of observations, $y_i \in \mathbb{R}$ and $\mathbf{x}_i \in \mathbb{R}^m$ ($i = 1, ..., n$) are response and explanatory variables, respectively, and $\varepsilon_i$ ($i = 1, ..., n$) is a random variable with mean 0. Moreover, $f \in \mathbb{F}$, where $\mathbb{F}$ is the set of all monotonic concave functions. Hence the problem is:

$$\min_{\varepsilon, f} \frac{1}{2} \sum_{i=1}^{n} \varepsilon_i^2$$
$$s.t.$$
$$y_i = f(\mathbf{x}_i) + \varepsilon_i, \quad i = 1, ..., n,$$
$$f \in \mathbb{F}.$$

---

[2] MCLS function is also known as convex nonparametric least squares (CNLS).

The set $\mathbb{F}$ is infinite-dimensional. However, it is shown that a piecewise linear function generates the best fit (Hildreth, 1954; Kuosmanen, 2008), and it is estimated by the following problem

$$\min_{\varepsilon,\alpha,\beta} \frac{1}{2}\sum_{i=1}^{n} \varepsilon_i^2$$
$$s.t. \qquad (1)$$
$$y_i = \alpha_i + \mathbf{x}_i \boldsymbol{\beta}_i' + \varepsilon_i, \quad i = 1,\dots n,$$
$$\alpha_i + \mathbf{x}_i \boldsymbol{\beta}_i' \le \alpha_j + \mathbf{x}_i \boldsymbol{\beta}_j', \quad i,j = 1,\dots,n,$$
$$\boldsymbol{\beta}_i \ge \mathbf{0}, \quad i = 1,\dots,n,$$

where $\mathbf{x}_i$ and $\boldsymbol{\beta}_i$ are the $i$-th rows of $\mathbf{X}$ and $\mathbf{B}$, respectively:

$$\mathbf{X} = (x_{ip})_{n\times m} = \begin{pmatrix} x_{11} & x_{12} & \dots & x_{1m} \\ x_{21} & x_{22} & \dots & x_{2m} \\ & & \ddots & \\ x_{n1} & x_{n2} & \dots & x_{nm} \end{pmatrix}, \mathbf{B} = (\beta_{ip})_{n\times m} = \begin{pmatrix} \beta_{11} & \beta_{12} & \dots & \beta_{1m} \\ \beta_{21} & \beta_{22} & \dots & \beta_{2m} \\ & & \ddots & \\ \beta_{n1} & \beta_{n2} & \dots & \beta_{nm} \end{pmatrix},$$

In problem (1), the first constraint specifies a hyperplane for every observation, with intercept $\alpha_i$ and slope variable $\boldsymbol{\beta}_i$. The second and the third constraints enforces concavity and monotonicity, respectively. Problem (1) is a basis to obtain variants of shape restricted least squares. For example, a CLS function can be estimated by problem (1) when the non-negativity constraint is relaxed (see sub-section 2.4). Figure 1 depicts examples of estimated CLS and MCLS functions.[3]

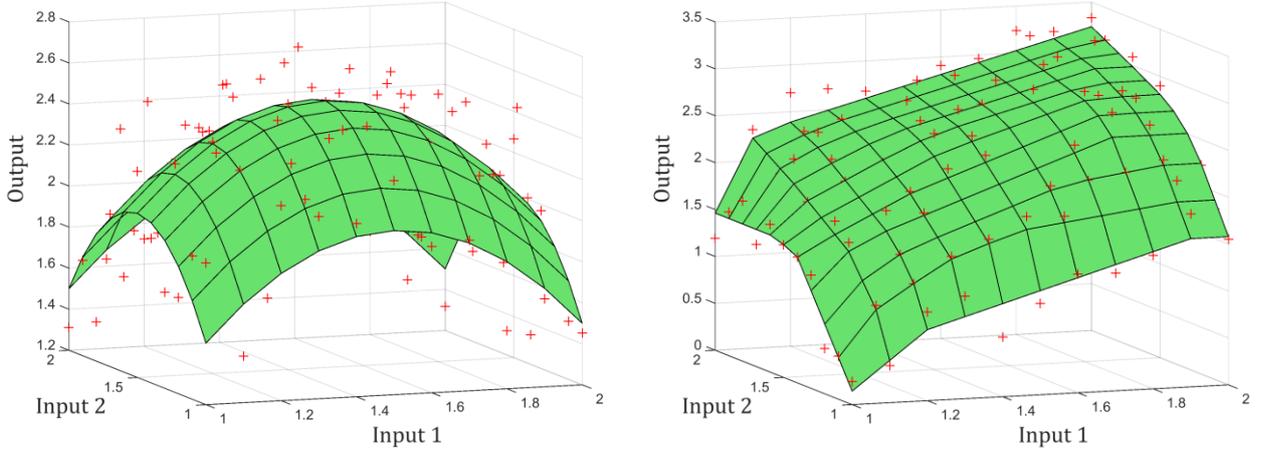

**Figure 1. Examples of two-variate shape restricted least squares. Left panel: CLS estimator, right panel: MCLS estimator.**

### 2.2 Penalization method

There are $O(n^2)$ constraints in problem (1), and to the best of our knowledge this is the main reason that solving (1) is very time consuming (Hannah & Dunson, 2013; Lee et al., 2013). Our approach to handle the large number of constraints is to eliminate all expect the non-negativities, and to penalize the objective function with constraints' violations. The penalty method consists of three steps: a) using the equality constraints to eliminate the intercept variables ($\alpha_i$), b) transforming the remaining constraints into equalities by adding slack variables, c) penalizing the objective function by the quadratic violations of the equality constraints. The reformulated problem is a penalized QP with only non-negativity constraints and it is solved quicker than the original problem (see Section 3). This problem is solved efficiently with off-the-shelf solvers (such as CPLEX and Mosek) and there is no need to develop a customized solver.

---
[3] For the sake of simplicity of the figure, we generated the data over a grid in the inputs space.

To start, consider that the intercept variables $\alpha_i$ ($i = 1, \ldots, n$) are eliminated by combining $\alpha_i = y_i - \mathbf{x}_i \boldsymbol{\beta}_i' - \varepsilon_i$ and the second constraint. By adding the slack variables $\mathbf{s}_i = (s_{i1}, \ldots, s_{in})'$, this problem is written as:

$$\min_{\varepsilon, \beta, s} \frac{1}{2} \sum_{i=1}^{n} \varepsilon_i^2$$
$$s.t. \qquad (2)$$
$$y_j - y_i = \varepsilon_j - \varepsilon_i + (\mathbf{x}_j - \mathbf{x}_i)\boldsymbol{\beta}_j' + s_{ij}, \quad i, j = 1, \ldots, n,$$
$$s_{ij} \geq 0, \boldsymbol{\beta}_i \geq \mathbf{0}, \quad i = 1, \ldots, n.$$

The algebraic calculations of the penalty term is simplified if the problem is presented in a succinct matrix form. To this end, a vector of variables is defined as:

$$\boldsymbol{\psi} = (\boldsymbol{\beta}^1, \boldsymbol{\beta}^2, \ldots, \boldsymbol{\beta}^m, \mathbf{s}_1, \mathbf{s}_2, \ldots, \mathbf{s}_n)',$$

where the slopes and slacks are stacked together to form a vector of size $mn + n^2$. Hereafter the superscript $p$ denotes the $p$-th column of matrices. Auxiliary matrices $\mathbf{E}_i$, $\boldsymbol{\mathcal{X}}_i$, and invertible $\mathbf{A}_i$ are defined in Appendix 1 in such a way that the second constraint of problem (2) is $(\mathbf{I} - \mathbf{E}_i)\mathbf{y} = \mathbf{A}_i \boldsymbol{\varepsilon} + \boldsymbol{\mathcal{X}}_i \boldsymbol{\psi}$ ($i = 1, \ldots, n$), where $\mathbf{I}$ is the identity matrix of order $n$. Thus $\boldsymbol{\varepsilon}$ is calculated by the following equation for any $i = 1, \ldots, n$:

$$\boldsymbol{\varepsilon} = -\mathbf{A}_i^{-1} \boldsymbol{\mathcal{X}}_i \boldsymbol{\psi} + \mathbf{A}_i^{-1}(\mathbf{I} - \mathbf{E}_i)\mathbf{y}. \qquad (3)$$

Theorem 1 calculates the objective function by using $i = 1$.

**Theorem 1.** $\boldsymbol{\varepsilon}'\boldsymbol{\varepsilon} = \boldsymbol{\psi}' \boldsymbol{\mathcal{X}}_1' \mathbf{A}_1^{-1'} \mathbf{A}_1^{-1} \boldsymbol{\mathcal{X}}_1 \boldsymbol{\psi} - 2\mathbf{y}' \left(\mathbf{I} - \frac{1}{n}\mathbf{1}\right) \mathbf{A}_1^{-1} \boldsymbol{\mathcal{X}}_1 \boldsymbol{\psi} + \mathbf{y}' \left(\mathbf{I} - \frac{1}{n}\mathbf{1}\right) \mathbf{y}$, where $\mathbf{1}$ is a $n \times n$ matrix of ones.

**Proof.** See Appendix 2.

In a similar approach, Theorem 2 calculates a penalty term as the sum of the quadratic violations of the constraints of (2).

**Theorem 2.** $\sum_{i=1}^{n} \sum_{j=1}^{n} \left( (y_j - y_i) - (\varepsilon_j - \varepsilon_i + (\mathbf{x}_j - \mathbf{x}_i)\boldsymbol{\beta}_j' + s_{ij}) \right)^2 =$

$$\boldsymbol{\psi}'(\sum_{i=2}^{n}(\boldsymbol{\mathcal{X}}_1 - \mathbf{A}_1 \mathbf{A}_i^{-1} \boldsymbol{\mathcal{X}}_i)'(\boldsymbol{\mathcal{X}}_1 - \mathbf{A}_1 \mathbf{A}_i^{-1} \boldsymbol{\mathcal{X}}_i))\boldsymbol{\psi}.$$

**Proof.** See Appendix 2.

Combining the results of theorems, *penalized MCLS* is defined as:

$$\min_{\boldsymbol{\psi}} \frac{1}{2} \boldsymbol{\psi}' \mathbf{H} \boldsymbol{\psi} + \mathbf{c}\boldsymbol{\psi}$$
$$s.t. \qquad (4)$$
$$\boldsymbol{\psi} \geq \mathbf{0},$$

where $\mathbf{H} = \mathbf{Q} + M^2\mathbf{V}$, $M$ is a large positive number, and there are

$$\mathbf{Q} = \mathcal{X}_1'\mathbf{A}_1^{-1'}\mathbf{A}_1^{-1}\mathcal{X}_1,$$
$$\mathbf{V} = \sum_{i=2}^n (\mathcal{X}_1 - \mathbf{A}_1\mathbf{A}_i^{-1}\mathcal{X}_i)'(\mathcal{X}_1 - \mathbf{A}_1\mathbf{A}_i^{-1}\mathcal{X}_i),$$
$$\mathbf{c} = -\mathbf{y}'\left(\mathbf{I} - \frac{1}{n}\mathbf{1}\right)\mathbf{A}_1^{-1}\mathcal{X}_1,$$
$$\gamma = \mathbf{y}'\left(\mathbf{I} - \frac{1}{n}\mathbf{1}\right)\mathbf{y}.$$

The objective of problem (4) is not the sum of squared errors (SSR) of the original problem. SSR is computed by Theorem 1, or simply by the following equation:

$$\sum_{i=1}^n \hat{\varepsilon}_i^2 = \boldsymbol{\psi}^{*\prime}\mathbf{Q}\boldsymbol{\psi}^* + 2\mathbf{c}\boldsymbol{\psi}^* + \gamma, \tag{5}$$

where $\hat{\varepsilon}$ is the error and $\boldsymbol{\psi}^*$ is the optimal solution to (4).

The solution to (4) converges to the optimal solution of (2) when $M$ increases to infinity. The conventional algorithm for choosing a sufficiently large $M$ is to start by an initial value and iteratively increasing it until the convergence is satisfactory. This result is shown in Theorem 3 below. In applications, the magnitude of penalty should depend on the magnitude of the problem data. We may use a large enough $M$ such that $\boldsymbol{\psi}^{*\prime}\mathbf{V}\boldsymbol{\psi}^*$ be as close as possible to zero.

**Theorem 3.** The following properties hold:

a) Problem (4) has an optimal solution for any given $M \geq 0$,
b) Let $\boldsymbol{\psi}^*$ and $\boldsymbol{\psi}^*(M)$ be the optimal solutions to problems (2) and (4), respectively. Then $\boldsymbol{\psi}^*(M) \to \boldsymbol{\psi}^*$ as $M \to \infty$.

**Proof.** See Appendix 2.

As it is shown in Theorem 3, optimal solution to (4) is convergent to the optimal solution to (2), and theoretically, a very large $M$ does not cause an issue in the convergence. However, a very large value of $M$ may cause numerical instability, which is mainly due to rounding errors. In practice, the value of the penalty term in (4) tends to zero if a very big $M$ is used, and thus the solution to (4) is in a tight neighborhood of feasible region of (2). The behavior of big $M$ and a rule for choosing a proper $M$ are explained in Appendix 2.

Penalized MCLS (4) is a non-negative unconstrained QP, which is solved by available QP solvers. Matrices $\mathbf{Q}$, $\mathbf{V}$, and vector $\mathbf{c}$ are sparse and readily computable. A number of simplifications for the calculations are discussed in Appendix 1.

### 2.3 Dual of penalized MCLS

The matrices in the dual of problem (4) are less dense, and this sensibly reduces the solution time. It is straightforward to obtain the Lagrangian *dual of penalized MCLS* as follows (see Appendix 2):

$$\min \frac{1}{2}\mathbf{w}'\mathbf{w}$$
$$s.t. \tag{6}$$
$$\mathbf{F}'\mathbf{w} + \mathbf{c}' \geq \mathbf{0},$$

where **w** is a $n^2$-vector of dual variables. There is $\mathbf{H} = \mathbf{F}'\mathbf{F}$ and $\mathbf{F} = (\mathbf{A}_1^{-1}\mathcal{X}_1, M(\mathcal{X}_2 - \mathbf{A}_2\mathbf{A}_1^{-1}\mathcal{X}_1),$ ..., $M(\mathcal{X}_n - \mathbf{A}_n\mathbf{A}_1^{-1}\mathcal{X}_1))$. The estimated error term is computed by $\hat{\boldsymbol{\varepsilon}} = -\mathbf{w}^* + \left(\mathbf{I} - \frac{1}{n}\mathbf{1}\right)\mathbf{y}$, where $\mathbf{w}^*$ is the optimal solution to (6).

There are several advantages to use (6). This problem is separable and matrix **F** is sparse, hence it is expected to be solved faster than problem (4) (Vanderbei, 2001, sec. 23). The experimental results in Section 3 show the superior performance of problem (6). Moreover, building problem (4) starts by making **F** and then computing $\mathbf{F}'\mathbf{F}$ to form **H**. By avoiding this matrix multiplication, fewer computations are needed and one source of numerical errors is removed. Another advantage is that matrix **F** is less dense than matrix **H** and hence, the memory usage decreases.

Problem (6) can directly be used in Matlab and R in its current matrix form. However, mathematical programming software, such as GAMS, AIMMS and AMPL, use the summation forms. To be able to use such software we present the summation form of problem (6) below. The performance of the solver does not depend on the programming language or the format of the problem. In practice, the solution time of problem (6) and its summation form (problem 7) with the same solver are the same.[4] The error is estimated by $\hat{\varepsilon}_i = y_i - \bar{y} - w_{1i}/\sqrt{2}$, where $\bar{y}$ is the average of response variables.

$$\min \frac{1}{2}\sum_{i,j=1}^n w_{ij}^2$$
$$s.t. \qquad\qquad\qquad\qquad\qquad\qquad\qquad\qquad\qquad (7)$$
$$\sum_{i\geq 2}(x_{1p} - x_{ip})\sum_{j\geq 2} w_{ij} \geq 0, p = 1, ..., m,$$
$$\frac{\sqrt{2}}{n}(x_{1p} - x_{ip})\left(\sum_j w_{1j} - nw_{1i}\right) - M\sum_{j\geq 2}(x_{1p} - x_{jp})w_{ji} + 2(y_i - \bar{y})(x_{1p} - x_{ip}) \geq 0, p = 1, ..., m, i = 2, ..., n,$$
$$\frac{\sqrt{2}}{n}\sum_i w_{1i} + M\sum_{i\geq 2} w_{i1} \geq 0,$$
$$-\frac{\sqrt{2}}{n}\left(\sum_j w_{1j} + nw_{1i}\right) + M\sum_{j\geq 2} w_{ji} - 2(y_i - \bar{y}) \geq 0, i = 2, ..., n,$$
$$\sum_{j\geq 2} w_{ij} \geq 0, i = 2, ..., n,$$
$$w_{ij} \leq 0, i, j = 2, ..., n, i \neq j,$$
$$w_{i1} \leq 0, i = 2, ..., n.$$

In order to increase the speed of the problem, the relaxation method proposed by Lee et al. (2013) can be combined with the reformulated problem proposed in this paper. Using this method[5], the second constraint of problem (1) is relaxed for some pairs of $(i, j)$. Let $I$ be the set of pairs $(i, j)$ of the seconds constraint of problem (1), which is determined in an iteration of the method of Lee et al. (2013). Let $K_I = \{k: k > nm \text{ and } k \neq (nm + (i-1)n + j) \text{ and } (i,j) \in I\}$ as a subset of $\{1, ..., nm + n^2\}$. The QP of Lee et al. (2013) can be replaced by problem (4) when $\psi_k = 0, k \in K_I$. Similarly, problem (6) can be used by removing row $k \in K_I$ of $\mathbf{F}'\mathbf{w} + \mathbf{c}' \geq \mathbf{0}$.

### 2.4 Variants of shape restricted least squares
A concave function is estimated by problem (1) if slope variables are free of sign, i.e. by removing $\beta_i \geq 0$ $(i = 1, ..., n)$. Furthermore, if the direction of inequalities is reversed, this problem estimates a convex least squares function. The reformulated problem (problem 4) and its dual (problem 6) can be easily adapted to solve different variants of the shape restricted least squares.

---
[4] The generation times of the problems may differ, and they are depended to the programming language.
[5] Method of Lee et al. (2013) is implemented in Ray, Kumbhakar, & Dua (2015).

In this sub-section, problems (4) and (6) are adapted for estimating a CLS function. The difference between CLS and MCLS is that the regression hyperplanes of CLS, which are defined by the first constraint of problem (1), are free of sign. This is shown by $\boldsymbol{\psi}_i \geq \boldsymbol{0}$ ($i > mn$) in *penalized CLS* problem as follows:

$$\min \frac{1}{2} \boldsymbol{\psi}' \mathbf{H} \boldsymbol{\psi} + \mathbf{c} \boldsymbol{\psi}$$
$$s.t.$$
$$\boldsymbol{\psi}_i \geq 0, \ j > mn,$$

where $\mathbf{H}$ and $\mathbf{c}$ are the same as in problem (4). The *dual of penalized CLS* is also developed similarly:

$$\min \frac{1}{2} \mathbf{w}' \mathbf{w}$$
$$s.t. \qquad\qquad\qquad\qquad (8)$$
$$\mathbf{F}' \mathbf{w} + \mathbf{c}' = \boldsymbol{\mu},$$
$$\boldsymbol{\mu}_j \geq 0, \ j > mn.$$

Adapting problems (4) and (6) to solve a convex problem is straightforward and it is done by reversing the signs of the slack variables of problem (1).

## 3. Experimental results

In this section, an experiment is designed to analyze the computational performance of (1), (4), and (6). The aim is to benchmark the penalization method (problems 4 and 6) against the original problem. The problems are solved on a laptop computer running Windows with an Intel Core i5 CPU 2.6 GHz and 8 gigabytes of RAM. The methods are implemented in Matlab, and Mosek 7 solves the QP problems. The measure of performance in this experiment is the average mean squared error (AMSE):

$$AMSE^\omega = \frac{1}{T} \sum_{t=1}^{T} \sqrt{\frac{SSR_t^\omega}{n}}, \ SSR_t^\omega = \sum_{i=1}^{N}(y_i - y_i^\omega)^2,$$

where $T$ is the number of simulations, $n$ is the number of observations, and $\omega \in$ {MCLS, PMCLS, dual of PMCLS} where

> MCLS: Problem (1), the original formulation of monotonic concave least squares,
> PMCLS: Penalized MCLS, problem (4),
> Dual of PMCLS: Dual of penalized MCLS, problem (6).

A measure for the goodness of fit of (1) is the coefficient of determination, $R^2$, which is one minus the ratio of SSR to the total sum of squares: $SST = \sum_{i=1}^{n}(y_i - \bar{y})^2$, where $\bar{y}$ is the mean value of $y$. Let $R_4^2$ be the coefficient of determination that is computed from the optimal solution to problem (4), and $\rho = 1 - (\boldsymbol{\psi}^{*'} \mathbf{H} \boldsymbol{\psi}^* + 2\mathbf{c}\boldsymbol{\psi}^* + \gamma)/SST$. In case of no violation, there is $R^2 = R_4^2 = \rho$. Since problem (2) allows for violations, $R_4^2$ is an upper bound for $R^2$ ($R^2 \leq R_4^2$). In case there are some violations, $R_4^2 = \rho + M^2 R_{aug}^2$ where $R_{aug}^2 = \boldsymbol{\psi}^{*'} \mathbf{V} \boldsymbol{\psi}^* / SST$. If $R_{aug}^2 \to 0$ then $\rho \to R^2$. $R_{aug}^2$ is the amount by which $R^2$ is augmented when some violations exist. It is unit free (like $R^2$) and it is used to measure the accuracy of (4) and (6). It is computed even if the solution to problem (1) is not calculated. $R_{aug}^2$ is desired to be close to zero.

Similar to Lee et al. (2012), data is generated from the function $y = \Pi_{r=1}^{m} x_r^{0.5/m} + \varepsilon$, where inputs are drawn from a uniform distribution in the interval $[10,100]$. The error term is generated from a normal distribution with mean zero and standard deviation of 10. There are totally 98 scenarios, with different number of observations from 50 to 700, and different number of inputs from 2 to 8. Each scenario is simulated 20 times to compute the average solution time. Figure 1 compares the solution times for solving the problems with four inputs using three formulations presented in this paper.

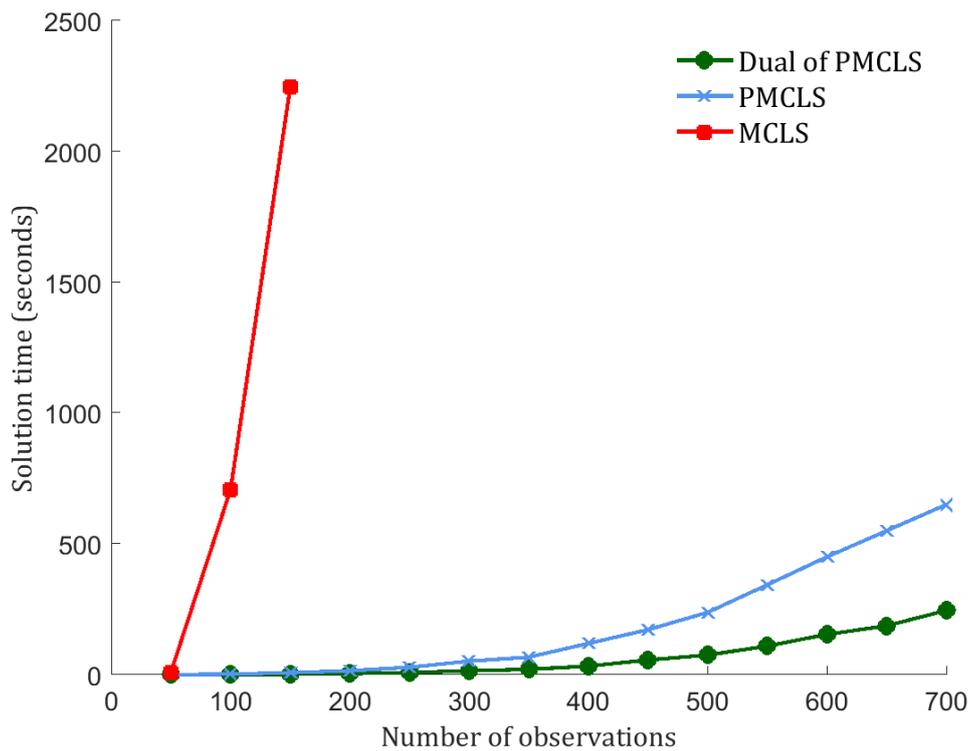

**Figure 2. Comparison of solution times for the scenario with four inputs. MCLS cannot be solved for problems with more than 150 observations via (1).**

As Figure 2 depicts, a problem with 4 inputs and more than 100 observations cannot be solved via (1), while it is solved in a relatively short time via (4) and (6). For example, a problem with 700 observations is solved in around four minutes via (4) and in around eleven minutes via (6). The solution times are presented in Table 1. Both (4) and (6) solve larger problems in a manageable time. As we expected from the discussion in sub-section 2.4, the dual problem is solved significantly faster.

**Table 1. The solution times in seconds for dual of PMCLS (D. PMCLS), PMCLS and MCLS. The standard deviations are in parenthesis. Maximum running time of solver is one hour.**

| Inputs | 2 | | | 3 | | | 4 | | | 5 | | | 6 | | | 7 | | | 8 | | |
|---|---|---|---|---|---|---|---|---|---|---|---|---|---|---|---|---|---|---|---|---|---|
| Obs. | D.PMCLS | PMCLS | MCLS | D.PMCLS | PMCLS | MCLS | D.PMCLS | PMCLS | MCLS | D.PMCLS | PMCLS | MCLS | D.PMCLS | PMCLS | MCLS | D.PMCLS | PMCLS | MCLS | D.PMCLS | PMCLS | MCLS |
| 50 | 0.1 (0) | 0.3 (0.1) | 9.1 (1.6) | 0.2 (0.1) | 0.5 (0.3) | 9.2 (1.6) | 0.5 (0.2) | 0.7 (0.4) | 9.4 (0.6) | 0.5 (0.3) | 0.8 (0.4) | 10.5 (2.6) | 0.6 (0.2) | 0.9 (0.4) | 11.0 (3.5) | 0.6 (0.3) | 1.2 (0.4) | 13.2 (2.9) | 0.6 (0.2) | 1.3 (0.2) | 15.0 (2.7) |
| 100 | 0.7 (0.1) | 1.1 (0.1) | 1.6 (0.5) | 0.7 (0.2) | 1.4 (0.6) | 704.4 (45.4) | 0.9 (0.4) | 2.2 (1.1) | 707.9 (49) | 1.0 (0.7) | 3.0 (1.6) | 958.8 (66.7) | 1.1 (0.5) | 3.1 (0.8) | 1163.0 (94.9) | 1.1 (0.5) | 4.4 (1.6) | | 1.5 (0.6) | 6.1 (1.5) | |
| 150 | 1.5 (0.3) | 2.8 (0.7) | 3.0 (0.7) | 1.6 (0.9) | 3.9 (1) | 1928.4 (96.4) | 2.0 (0.5) | 8.0 (2.5) | 2245.6 (113.4) | 3.9 (1.8) | 9.0 (3) | | 4.0 (1.4) | 11.9 (2.8) | | 5.4 (2.5) | 14.1 (2.8) | | 5.7 (1.2) | 18.8 (2.7) | |
| 200 | 2.3 (1.1) | 7.4 (1.5) | 8.2 (0.9) | 2.9 (1.3) | 11.7 (1.7) | 3532.5 (187.6) | 4.4 (0.8) | 15.4 (2.2) | | 5.1 (1.8) | 16.1 (3.4) | | 8.4 (3.1) | 26.8 (5.1) | | 8.9 (2.2) | 33.1 (3.8) | | 10.5 (3.4) | 48.6 (5.4) | |
| 250 | 4.3 (1.1) | 15.2 (3.4) | 21.1 (6.7) | 5.4 (2.1) | 23.3 (2.1) | | 8.9 (1.2) | 28.5 (7.9) | | 12.8 (2.8) | 39.2 (7.2) | | 13.8 (8) | 52.9 (8.4) | | 15.7 (5.1) | 65.5 (4.6) | | 17.4 (4.1) | 98.2 (8.3) | |
| 300 | 9.2 (2.5) | 25.1 (5.8) | 32.5 (6.9) | 10.5 (1.7) | 36.0 (9.5) | | 14.8 (3.1) | 52.0 (4.9) | | 15.6 (4.7) | 57.6 (9.3) | | 21.0 (3) | 91.2 (10.9) | | 29.7 (7.4) | 141.5 (14.1) | | 29.8 (9.8) | 202.3 (15.5) | |
| 350 | 16.2 (3.1) | 39.9 (6.4) | 55.1 (1.1) | 19.1 (5.2) | 53.9 (13.7) | | 20.8 (3) | 67.3 (5.5) | | 30.0 (11.3) | 107.9 (18.5) | | 44.5 (14.7) | 175.9 (5.9) | | 56.9 (14.9) | 210.5 (10.9) | | 56.4 (13.8) | 293.6 (15.6) | |
| 400 | 23.5 (3.7) | 54.6 (5.6) | 94.5 (8.5) | 31.2 (3.1) | 96.8 (12.2) | | 33.0 (5.2) | 119.7 (31.7) | | 48.1 (6.1) | 182.4 (19) | | 65.5 (20.5) | 257.3 (28.5) | | 77.9 (23.8) | 353.9 (41.5) | | 104.7 (26.9) | 472.7 (33.7) | |
| 450 | 32.8 (4.8) | 90.1 (11.9) | 157.9 (8.9) | 48.1 (4.4) | 134.1 (10.3) | | 56.5 (7.9) | 172.2 (25) | | 64.8 (9.5) | 277.8 (26.2) | | 83.2 (8) | 356.9 (31.3) | | 83.2 (0.8) | 464.8 (38.9) | | 120.3 (28.5) | 678.8 (83) | |
| 500 | 49.1 (5.4) | 141.1 (13.2) | 251.6 (17.2) | 60.0 (7.8) | 206.9 (33.2) | | 75.3 (11.8) | 238.3 (19.2) | | 105.2 (14.1) | 339.0 (47.8) | | 120.9 (14) | 466.4 (26.2) | | 139.7 (16.5) | 684.9 (50.9) | | 179.9 (67.8) | 781.6 (3.3) | |
| 550 | 75.8 (4) | 203.8 (31.6) | 365.4 (20.6) | 81.1 (11.1) | 270.2 (41.9) | | 109.7 (13.2) | 343.6 (19.4) | | 132.4 (11.2) | 502.2 (66.7) | | 195.0 (62.9) | 626.2 (55.6) | | 194.4 (13.8) | 949.8 (83.7) | | 256.4 (87.4) | 981.4 (125.4) | |
| 600 | 88.3 (22.8) | 244.8 (30) | 479.6 (39.9) | 122.6 (13.6) | 338.9 (10.7) | | 154.2 (23.9) | 450.7 (74.5) | | 183.4 (52.2) | 627.3 (50) | | 244.8 (7.9) | 1021.5 (96.9) | | 330.0 (97.5) | 1114.1 (96.4) | | 422.2 (147.6) | 1629.0 (217.9) | |
| 650 | 123.7 (25.4) | 341.4 (46.3) | 665.8 (28.4) | 143.3 (16.4) | 466.1 (13) | | 187.0 (39.5) | 550.1 (66.1) | | 226.4 (12.8) | 727.7 (22.2) | | 339.1 (93.9) | 1132.4 (43.3) | | 447.5 (87.2) | 1372.9 (373.3) | | 485.1 (99.6) | 1990.8 (290.6) | |
| 700 | 144.2 (12.1) | 402.0 (38.3) | 867.0 (51.1) | 220.6 (12.9) | 550.9 (25.9) | | 245.7 (22.4) | 650.5 (94.6) | | 316.9 (42.6) | 948.2 (91.9) | | 412.2 (45.9) | 1368.0 (85.4) | | 571.1 (41.4) | 1586.6 (677) | | 629.1 (10.6) | 2757.3 (323.8) | |

The accuracy of the solutions is assessed by the AMSE statistic and $R^2_{aug}$ on Table 2. The results indicate that the methods have similar performances, while dual of PMCLS performs slightly better. The value of $R^2_{aug}$ for the dual of PMCLS and PMCLS in all scenarios is less than $10^{-7}$. Therefore, PMCLS and its dual obtain the optimal solution to MCLS in a reasonably shorter time. The dual problem is solved significantly faster than PMCLS, and its performance with regard to the AMSE statistic is slightly better, therefore the dual of PMCLS is the preferred formulation for solving a monotonic concave least squares problem.

**Table 2.** AMSE of the three formulations. The value of $R^2_{aug}$ for D.PMCLS and PMCLS in all scenarios is less than $10^{-7}$.

| Inputs | 2 | | | 3 | | | 4 | | | 5 | | | 6 | | | 7 | | | 8 | | |
|---|---|---|---|---|---|---|---|---|---|---|---|---|---|---|---|---|---|---|---|---|---|
| Obs. | D. PMCLS | PMCLS | MCLS | D. PMCLS | PMCLS | MCLS | D. PMCLS | PMCLS | MCLS | D. PMCLS | PMCLS | MCLS | D. PMCLS | PMCLS | MCLS | D. PMCLS | PMCLS | MCLS | D. PMCLS | PMCLS | MCLS |
| 50 | 8.78 | 8.79 | 8.79 | 7.68 | 7.71 | 7.72 | 7.19 | 7.19 | 7.19 | 6.18 | 6.23 | 6.23 | 6.19 | 6.21 | 6.22 | 5.95 | 6.00 | 6.00 | 7.27 | 7.30 | 7.30 |
| 100 | 9.55 | 9.55 | 9.62 | 8.44 | 8.45 | 8.45 | 8.31 | 8.31 | 8.31 | 7.22 | 7.22 | 7.23 | 7.08 | 7.12 | 7.14 | 7.58 | 7.60 | | 6.82 | 6.83 | |
| 150 | 9.99 | 9.99 | 9.99 | 8.60 | 8.60 | 8.60 | 8.69 | 8.69 | 8.70 | 7.48 | 7.49 | | 7.87 | 7.90 | | 7.88 | 7.92 | | 7.27 | 7.31 | |
| 200 | 9.47 | 9.48 | 9.48 | 9.20 | 9.20 | 9.20 | 8.70 | 8.71 | | 8.25 | 8.26 | | 7.81 | 7.83 | | 8.01 | 8.06 | | 7.94 | 7.96 | |
| 250 | 9.42 | 9.43 | 9.43 | 9.03 | 9.03 | | 8.99 | 9.00 | | 8.43 | 8.47 | | 8.24 | 8.27 | | 8.43 | 8.44 | | 7.72 | 7.74 | |
| 300 | 9.38 | 9.39 | 9.39 | 9.72 | 9.73 | | 8.97 | 8.97 | | 8.51 | 8.52 | | 8.36 | 8.39 | | 8.58 | 8.61 | | 8.73 | 8.76 | |
| 350 | 9.79 | 9.80 | 9.80 | 9.59 | 9.61 | | 9.15 | 9.16 | | 8.71 | 8.74 | | 8.86 | 8.88 | | 8.87 | 8.89 | | 8.30 | 8.33 | |
| 400 | 9.35 | 9.37 | 9.36 | 9.47 | 9.48 | | 9.41 | 9.42 | | 9.07 | 9.10 | | 8.91 | 8.95 | | 8.60 | 8.63 | | 8.17 | 8.23 | |
| 450 | 9.81 | 9.83 | 9.82 | 9.45 | 9.46 | | 9.17 | 9.19 | | 8.78 | 8.80 | | 8.94 | 8.97 | | 8.75 | 8.76 | | 8.46 | 8.49 | |
| 500 | 9.51 | 9.52 | 9.52 | 9.62 | 9.62 | | 9.30 | 9.32 | | 8.99 | 9.01 | | 8.47 | 8.49 | | 8.61 | 8.64 | | 8.76 | 8.77 | |
| 550 | 9.98 | 9.99 | 9.99 | 9.43 | 9.44 | | 9.28 | 9.29 | | 8.72 | 8.76 | | 9.03 | 9.05 | | 8.76 | 8.81 | | 8.76 | 8.78 | |
| 600 | 10.06 | 10.07 | 10.07 | 9.78 | 9.80 | | 9.39 | 9.41 | | 9.16 | 9.17 | | 9.02 | 9.05 | | 8.77 | 8.81 | | 8.85 | 8.89 | |
| 650 | 9.84 | 9.85 | 9.84 | 9.68 | 9.69 | | 9.48 | 9.50 | | 9.02 | 9.05 | | 9.14 | 9.17 | | 8.85 | 8.89 | | 8.81 | 8.87 | |
| 700 | 9.77 | 9.77 | 9.77 | 9.45 | 9.47 | | 9.20 | 9.23 | | 9.01 | 9.04 | | 8.94 | 8.98 | | 8.96 | 9.07 | | 9.96 | 10.00 | |

Combining the results of the experiment together, we conclude that PMCLS and its dual are computationally more efficient than the original formulation of the problem, and the dual of PMCLS has the best performance in running times.

Furthermore, the performance of the methods is tested on large problems by simulating scenarios with thousands of observations. The starting number of observations is 1000 with increment of 500, and the maximum running time is five hours for every simulation. Every scenario is simulated 5 times. The results are reported in Table 3. The largest problem that is solved via problem (6) has 2500 observations. The average solution time of the dual of PMCLS for problems with 2500 observations is around 187 minutes. The largest problem that problem (4) solved has 2000 observation and the average solution time is 264 minutes. The maximum $R^2_{aug}$ in all scenarios is $10^{-4}$. The original formulation (problem 1) cannot solve any of the large problems in the time limit.

**Table 3.** The solution times in seconds for problems with more than 1000 observations.

| | D. PMCLS | PMCLS | MCLS |
|---|---|---|---|
| **1000** | 914 (143) | 2204 (301) | - |
| **1500** | 2966 (423) | 8284 (792) | - |
| **2000** | 7258 (1226) | 15856 (1785) | - |
| **2500** | 11259 (1988) | - | - |

By using penalized MCLS and its dual multivariate concave and convex least squares functions can be estimated for problems with several thousands of observations. As we discussed in the introduction section, the computation time is one of the major difficulties in using shape restricted least squares, which can be effectively eliminated by using problems (4) and (6).

## 4. Empirical study

The computation advantages of problem (6) over problem (1) is used in this section to analyze the room rates of a sample of hotels. The data consists of the average room rates and 12 hotel attributes of 126 hotels with minimum star rating of 2 in Finland. The data set is collected from Expedia.com in January 2016. The conventional approach to explain the price based on the attributes of the hotel is to use the hedonic pricing analysis (see for example Chen & Rothschild, 2010; Espinet, Saez, Coenders, & Fluvia, 2003; Rigall-I-Torrent & Fluvià, 2011; Semere, 2014; Thrane, 2007). In this section, we first show that a monotonic concave function generates a better fit than a linear function. This is not however a surprise because set $\mathcal{F}$ includes linear functions and a linear hedonic function is a restricted case of a piecewise linear function. Secondly, we estimate an efficient frontier that shows the maximum of room rate for every hotel if their pricing strategy is efficient in comparison with the hotels in the sample. Hotel attributes are explained in Table 4. The selection of attributes is based on similar researches.

**Table 4. Descriptions of attributes for the hotels in the sample ($n = 126$).**

| Attribute | Description | Mean | Std. Dev. |
|---|---|---|---|
| Average rate | Average room rate, logged | 5.14 | 0.38 |
| Rooms | Number of rooms | 162.33 | 96.24 |
| Stars | Star rating of hotel | 3.64 | 0.64 |
| Distance | Distance to central railway station (KM) | 2.53 | 3.69 |
| Chain | Hotel is associated with a chain (binary) | 0.75 | 0.44 |
| Breakfast | Free breakfast is included (binary) | 0.56 | 0.5 |
| Business | Business facilities, such as meeting rooms, are available (binary) | 0.65 | 0.48 |
| Fitness | Fitness facilities is present at hotel (binary) | 0.66 | 0.48 |
| Parking | Free parking is included (binary) | 0.21 | 0.41 |
| Hair dryer | Hair dryer is present at hotel (binary) | 0.75 | 0.44 |
| Pool | Pool is present at hotel (binary) | 0.45 | 0.5 |
| Sauna | Sauna is present at hotel (binary) | 0.82 | 0.39 |
| Spa | Spa is present as hotel (binary) | 0.87 | 0.34 |

Let $\mathbf{R}_i$ be a $m$ vector of attributes of hotel $i$, $p(\mathbf{R}_i)$ be the room rate, and $y_i$ be the logarithm of the average room rate. The natural logarithm of the average room rates are used to maintain a more nearly linear function (Gelman & Hill, 2007). Thus $y_i = p(\mathbf{R}_i) + \varepsilon_i$ where $\varepsilon_i$ is an error term. While the dual of PMCLS (problem 6) obtains the optimal solution in around 2 seconds, problem (1) cannot be solved for this sample of hotels within a one-hour limit ($n = 126, m = 12$).

The linear hedonic function is also estimated and the $R^2$ coefficients of both models are reported in Table 5. As it is shown, the estimated monotonic concave function generates a better fit. The $R^2$ of MCLS is 66% which is 1.8 times of $R^2$ of a linear function. Therefore, it is concluded that a monotonic piecewise linear function explains more of the variations of the room rates. However, problem (1) cannot be practically used to estimate such a piecewise linear function due to the lack of memory in the computer and to the lengthy solution times.

**Table 5. The coefficients of determination of the estimated linear and monotonic piecewise linear functions**

| Function | Linear | Monotonic piecewise linear |
|---|---|---|
| $R^2$ | 36% | 66% |

Let $\hat{p}(\mathbf{R}_i)$ be the estimated price for hotel $i$ via problem (6). The variation of $y_i$ from $\hat{p}(\mathbf{R}_i)$ may be either due to the performance of the pricing strategy of the hotel or it may be because of a random noise. The method of StoNED provides a basis to estimate the performance of the pricing strategy of the hotel. This method is well studied in several publications, for example see Keshvari and Kuosmanen, (2013) and Kuosmanen and Kortelainen (2012). By using StoNED, the error term ($\varepsilon$) is decomposed into a noise part ($v$) and an asymmetric inefficiency part ($u > 0$) such that $\varepsilon_i = u_i - v_i$. The inefficiency term $u_i$ is assumed to be half-normally distributed with the variance $\sigma_u^2$, and the noise term $v_i$ is assumed to be normally distributed with the zero mean and the variance $\sigma_v^2$. By using the method of moments the standard deviation of the inefficiency term is estimated as $\hat{\sigma}_u = \sqrt[3]{\widehat{M}_3 / \left[ \sqrt{2/\pi} \, (4/\pi - 1) \right]}$ where $\widehat{M}_3 = \sum_{i=1}^{n} \varepsilon_i^3 / n$ is the estimated third central moment of $\varepsilon$. The expected room rate in then estimated by $\hat{y}_i = \hat{p}(\mathbf{R}_i) + \hat{\sigma}_u \sqrt{2/\pi}$. The difference between the room rate and the estimated room rate via StoNED shows the amount by which the hotel can adjust the room rate to be efficient in the sample.

Figure 3 shows histogram of the estimated adjustments of room rate, i.e. $y_i - \hat{y}_i$. According to this figure, 45 hotels have a negative adjustment value, which means that based on this sample of hotels, there are some potentials for them to increase their room rate. Most of these hotels (34) may increase the prices by up to 50 euros, and two of the hotels have the adjustment value of 150 euros. On the other hand, there are 14 hotels with a positive adjustment value. It is suggested that 9 hotels decrease the prices by up to 50 euros and there are also some hotels that may decrease their room rates by up to 250 or 300 euros. This analysis of the room rate is based on the frontier analysis and further research is required to investigate and analyze the drivers of the room rates of the hotels in the sample. However, as Table 5 shows, this method gives a better fit than a traditional hedonic pricing analysis which is currently the main method for explaining the relations between the room rate and hotel attributes.

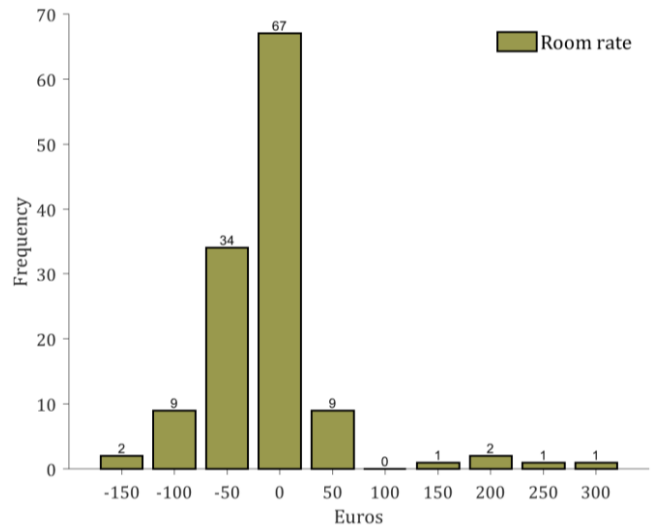

**Figure 3. Histogram of the room rates and the estimated room rates.**

## 5. Conclusions

In this paper, we proposed an alternative formulation for concave and convex regression via least squares. Despite the interesting properties and wide range of applications of such estimators, solving the problem is very time consuming. One major source of complexity is the number of constraints in the QP problems. In our proposal, we reformulate the problem as a non-negative unconstrained QP. This problem can be solved by using available QP solvers. The numerical tests show that our penalized monotonic concave regression and its dual perform significantly better than the original problem. To analyze the pricing strategies of a sample of 126 hotels in Finland, we estimate a piecewise linear concave function. While this problem cannot be solved via the original formulation of MCLS in one hour, it is solved in around two seconds via the dual of PMCLS. The results show that MCLS is a better fit than a linear function. The Matlab and R codes for solving the reformulated problems are also provided in this paper.

## Acknowledgements

The author would like to thank Professor Timo Kuosmanen (Aalto University) and Professor Andrew L. Johnson (Texas A&M University) for the helpful discussions and comments.

## Appendix 1. Developing matrix form of MCLS

Note that problem (2) has $n^2 - n$ constraints. By adding the condition $\sum_{i=1}^{n} \varepsilon_i = 0$ to problem (2) and repeating it $n$ times, there are $n^2$ constraints. The condition $\sum_{i=1}^{n} \varepsilon_i = 0$ is one of the finite sample properties of shape restricted regression estimator (see for example Seijo & Sen, 2011). Here we present an alternative proof to this property in Proposition 1.

**Proposition 1:** In the optimal solution to problem (1), the sum of residuals is zero.

**Proof.** See the Appendix 2.

Having this property, we categorize the constraints in (2) into $n$ blocks as follows:

Block $i$ $(i = 1, \ldots, n)$:
$$y_j - y_i = \varepsilon_j - \varepsilon_i + (x_j - x_i)\boldsymbol{\beta}_j + s_{ij}, \quad j = 1, \ldots, n, \; j \neq i,$$
$$\sum_{i=1}^{n} \varepsilon_i = 0,$$

where $\sum_{i=1}^{n} \varepsilon_i = 0$ is appended to all blocks such that there are $n$ constraints in every block. Here, our aim is to develop the matrix form of block $i$. To accomplish this purpose we use the following auxiliary matrices:

$$\mathbf{0}_{c \times d} = \text{matrix of zeros of size } c \times d,$$
$$\mathbf{1}_{c \times 1} = \text{vector of ones of size } c,$$
$$\mathbf{1} = \text{matrix of ones of size } n \times n,$$
$$\mathbf{I} = \text{the identity matrix of size } n,$$
$$\mathbf{E}_i = \begin{bmatrix} \mathbf{0}_{n \times (i-1)} & \mathbf{1}_{n \times 1} & \mathbf{0}_{n \times (n-i)} \end{bmatrix},$$
$$\mathbf{A}_i = \mathbf{I} - \mathbf{E}_i + \mathbf{E}'_i,$$
$$\boldsymbol{\mathcal{X}}_i = (\mathbf{d}_1, \ldots, \mathbf{d}_m, \mathbf{0}_{n \times n(i-1)}, \mathbf{I}, \mathbf{0}_{n \times n(n-i)}),$$
$$\mathbf{d}_p = diag(\mathbf{x}^p) - x_{ip}\mathbf{I}, \; p = 1, \ldots, m,$$

where $diag(x^p)$ refers to the diagonal matrix of vector $x^p$. $\boldsymbol{\mathcal{X}}_i$ is a sparse matrix of data, which is made of $m + n$ sub-matrices. The first $m$ sub-matrices ($\mathbf{d}_p$) are diagonal matrices. The next $n$ sub-matrices of $\boldsymbol{\mathcal{X}}_i$ consist of $n - 1$ zero matrices and one identity matrix. The identity matrix is placed such that $\boldsymbol{\mathcal{X}}_i \boldsymbol{\psi} = \sum_{p=1}^{m} (diag(x^p) - x_{ip}\mathbf{I})\boldsymbol{\beta}^p + \mathbf{s}_i$.

With the help of the auxiliary matrices, the $i$-th block of matrices is written as

$$(\mathbf{I} - \mathbf{E}_i)\mathbf{y} = \mathbf{A}_i \boldsymbol{\varepsilon} + \boldsymbol{\mathcal{X}}_i \boldsymbol{\psi}, \tag{A1}$$

where $\boldsymbol{\varepsilon} = (\varepsilon_1, \varepsilon_2, \ldots, \varepsilon_n)'$. Consider that constraint $\sum_{i=1}^{n} \varepsilon_i = 0$ is placed as the $i$-th equality in (A1). By using equation (A1), problem (2) is written as $\left\{\min_{\boldsymbol{\varepsilon}, \boldsymbol{\psi}} \frac{1}{2}\boldsymbol{\varepsilon}'\boldsymbol{\varepsilon} \; s.t. (\mathbf{I} - \mathbf{E}_i)\mathbf{y} = \mathbf{A}_i\boldsymbol{\varepsilon} + \boldsymbol{\mathcal{X}}_i\boldsymbol{\psi}, \; i = 1, \ldots, n, \boldsymbol{\psi} \geq \mathbf{0}\right\}$. Building the matrices $\mathbf{Q}$, $\mathbf{V}$, and the vector $\mathbf{c}$ are heavily based on the auxiliary matrices. To simplify the calculations, we summarize the necessary operations in the following proposition.

**Proposition 2.** Matrix $\mathbf{A}_i$ is invertible. Moreover, the following properties hold for the auxiliary matrices:

a) $\mathbf{A}_i^{-1} = \mathbf{I} - \frac{1}{n}\mathbf{1} + \frac{2}{n}\mathbf{E}_i - \mathbf{e}_{ii}$,

b) $\mathbf{A}_i^{-1}(\mathbf{I} - \mathbf{E}_i) = \mathbf{I} - \frac{1}{n}\mathbf{1}$,

c) $\mathbf{A}_i^{-1'}\mathbf{A}_i^{-1} = \mathbf{I} - \frac{1}{n}\mathbf{1} + \frac{1}{n}(\mathbf{E}_i + \mathbf{E}_i') - \mathbf{e}_{ii}$,

d) $\mathbf{A}_1\mathbf{A}_i^{-1} = \mathbf{I} - \mathbf{E}_1 - \mathbf{e}_{ii} + \mathbf{e}_{1i}$,

e) $\left(\mathbf{I} - \frac{1}{n}\mathbf{1}\right)\mathbf{A}_1^{-1} = \mathbf{I} - \frac{1}{n}\mathbf{1} + \frac{1}{n}\mathbf{E}_1 - \mathbf{e}_{11}$,

where $\mathbf{e}_{ij}$ is an $n \times n$ matrix whose $(i,j)$ element is 1 and other elements are zero.

**Proof.** See the Appendix 2.

Proposition 2 proves that matrix $\mathbf{A}_i$ is invertible. Hence, we calculate the closed form definition of the error vector $\boldsymbol{\varepsilon}$ as the following:

$$\boldsymbol{\varepsilon} = -\mathbf{A}_i^{-1}\mathbf{X}_i\boldsymbol{\psi} + \left(\mathbf{I} - \frac{1}{n}\mathbf{1}\right)\mathbf{y}, \quad i = 1, \dots, n. \tag{A2}$$

To estimate the residuals we use the first block of equation (A2), i.e. $i = 1$, but this choice is arbitrary and any of the blocks may be used. Equation (A2) is used in theorems 1 and 2 to obtain the penalty term.

Note that matrices $\mathbf{A}_1^{-1'}\mathbf{A}_1^{-1}$, $\mathbf{A}_1\mathbf{A}_i^{-1}$ and $\left(\mathbf{I} - \frac{1}{n}\mathbf{1}\right)\mathbf{A}_1^{-1}$ that are used in the calculations of the penalized problem, have closed form definitions in Proposition 2. These matrices depend only to the number of observations and not to the problem data. Hence, we may compute them prior to solve the problem. Such matrices may be used to enhance the speed of computations in large-scale problems.

# Appendix 2. Proofs and algebraic computations

**Proposition 1:** In the optimal solution to problem (1), the sum of residuals is zero.

**Proof.** To prove this proposition we use the optimality conditions of problem (1) based on the Karush–Kuhn–Tucker (KKT) conditions. Suppose $(\varepsilon_i^*, \alpha_i^*, \boldsymbol{\beta}_i^*), i = 1, \ldots, n$ is a global minimizer of problem (1). Hence, there exist parameters $\lambda_i, \mu_{ij}, \boldsymbol{\gamma}_i = (\gamma_{i1}, \ldots, \gamma_{im})'$ $(i, j = 1, \ldots, n)$ that satisfy the following conditions:

**KKT conditions for the CLS problem**

$$-2\varepsilon_i = \lambda_i, \quad i = 1, \ldots, n, \tag{A3}$$
$$\lambda_i + \sum_{j=1}^n \mu_{ij} - \sum_{j=1}^n \mu_{ji} = 0, \quad i = 1, \ldots, n, \tag{A4}$$
$$\mathbf{x}_i'\lambda_i + \sum_{j=1}^n \mathbf{x}_i'\mu_{ij} - \sum_{j=1}^n \mathbf{x}_i'\mu_{ji} + \boldsymbol{\gamma}_i = 0, \quad i = 1, \ldots, n,$$
$$\boldsymbol{\gamma}_i'\boldsymbol{\beta}_i = 0, \quad i = 1, \ldots, n,$$
$$\mu_{ij}(\alpha_i + \mathbf{x}_i\boldsymbol{\beta}_i - \alpha_j - \mathbf{x}_i\boldsymbol{\beta}_j) = 0, \quad i, j = 1, \ldots, n,$$
$$\boldsymbol{\gamma}_i \geq \mathbf{0}, \ \mu_{ij} \geq 0, \quad i, j = 1, \ldots, n,$$
$$y_i = \alpha_i + \mathbf{x}_i\boldsymbol{\beta}_i' + \varepsilon_i, \quad i = 1, \ldots n,$$
$$\alpha_i + \mathbf{x}_i\boldsymbol{\beta}_i' \leq \alpha_j + \mathbf{x}_i\boldsymbol{\beta}_j', \quad i, j = 1, \ldots, n,$$
$$\boldsymbol{\beta}_i \geq \mathbf{0}, \quad i = 1, \ldots, n,$$

where $\lambda_i$ and $\mu_{ij}$ are Lagrange multipliers of the first and the second constraints of the CLS problem, respectively, and $\boldsymbol{\gamma}_i$ is the vector of Lagrange multipliers of the nonnegativity constraints.

Among all the conditions above, we need (A3) and (A4). Consider that $\sum_{i=1}^n \sum_{j=1}^n \mu_{ij} = \sum_{i=1}^n \sum_{j=1}^n \mu_{ji}$. Hence, we have $\sum_{i=1}^n \lambda_i = 0$ from condition (A4). Consequently, there is $\sum_{i=1}^n \varepsilon_i = 0$ from condition (A3). ∎

**Proposition 2.** Matrix $\mathbf{A}_i$ is invertible. Moreover, the following properties hold for the auxiliary matrices in (6):

a) $\mathbf{A}_i^{-1} = \mathbf{I} - \frac{1}{n}\mathbf{1} + \frac{2}{n}\mathbf{E}_i - \mathbf{e}_{ii}$,
b) $\mathbf{A}_i^{-1}(\mathbf{I} - \mathbf{E}_i) = \mathbf{I} - \frac{1}{n}\mathbf{1}$,
c) $\mathbf{A}_i^{-1'}\mathbf{A}_i^{-1} = \mathbf{I} - \frac{1}{n}\mathbf{1} + \frac{1}{n}(\mathbf{E}_i + \mathbf{E}_i') - \mathbf{e}_{ii}$,
d) $\mathbf{A}_1\mathbf{A}_i^{-1} = \mathbf{I} - \mathbf{E}_1 - \mathbf{e}_{ii} + \mathbf{e}_{1i}$,
e) $\left(\mathbf{I} - \frac{1}{n}\mathbf{1}\right)\mathbf{A}_1^{-1} = \mathbf{I} - \frac{1}{n}\mathbf{1} + \frac{1}{n}\mathbf{E}_1 - \mathbf{e}_{11}$,

where $\mathbf{e}_{ij}$ is an $n \times n$ matrix which its $(i, j)$ element is 1 and other elements are zero.

**Proof.**

First, we show $\mathbf{A}_i$ is positive definite, and hence it is invertible. Let $\mathbf{q}$ be a nonzero vector of size $n$. There is $\mathbf{q}'\mathbf{A}_i\mathbf{q} = \mathbf{q}'(\mathbf{I} - \mathbf{E}_i + \mathbf{E}_i')\mathbf{q} = \mathbf{q}'\mathbf{q} > 0$.

To prove the other statements, we use the following equalities:

$\mathbf{1}\mathbf{E}_i = n\mathbf{E}_i$, $\mathbf{1}\mathbf{E}'_i = \mathbf{1}$, $\mathbf{1}\mathbf{e}_{ii} = \mathbf{E}_i$, $\mathbf{1}\mathbf{1} = n\mathbf{1}$, $\mathbf{E}_i\mathbf{E}_i = \mathbf{E}_i$, $\mathbf{E}_i\mathbf{E}'_i = \mathbf{1}$, $\mathbf{E}'_i\mathbf{E}_i = n\mathbf{e}_{ii}$, $\mathbf{e}_{ii}\mathbf{E}_i = \mathbf{e}_{ii}$, $\mathbf{E}_i\mathbf{e}_{ii} = \mathbf{E}_i$, $\mathbf{e}_{ii}\mathbf{e}_{ii} = \mathbf{e}_{ii}$,

and for $2 \leq i \leq n$: $\mathbf{E}_1\mathbf{E}_i = \mathbf{E}_i$, $\mathbf{E}'_i\mathbf{E}_1 = n\mathbf{e}_{i1}$, $\mathbf{E}_1\mathbf{e}_{ii} = \mathbf{0}$, $\mathbf{E}'_1\mathbf{e}_{ii} = \mathbf{e}_{1i}$.∎

**Theorem 1.** $\boldsymbol{\varepsilon}'\boldsymbol{\varepsilon} = \boldsymbol{\psi}'\mathcal{X}'_1\mathbf{A}_1^{-1'}\mathbf{A}_1^{-1}\mathcal{X}_1\boldsymbol{\psi} - 2\mathbf{y}'\left(\mathbf{I} - \frac{1}{n}\mathbf{1}\right)\mathbf{A}_1^{-1}\mathcal{X}_1\boldsymbol{\psi} + \mathbf{y}'\left(\mathbf{I} - \frac{1}{n}\mathbf{1}\right)\mathbf{y}$

**Proof.**

By using equation (6), the proof is straightforward. ∎

**Theorem 2.** $\sum_{i=1}^{n}\sum_{j=1}^{n}\left(\varepsilon_j - \varepsilon_i + (\mathbf{x}_j - \mathbf{x}_i)\boldsymbol{\beta}'_j + s_{ij} - (y_j - y_i)\right)^2 =$

$$\boldsymbol{\psi}'(\sum_{i=2}^{n}(\mathcal{X}_1 - \mathbf{A}_1\mathbf{A}_i^{-1}\mathcal{X}_i)'(\mathcal{X}_1 - \mathbf{A}_1\mathbf{A}_i^{-1}\mathcal{X}_i))\boldsymbol{\psi}.$$

**Proof.**

The error is calculated by equation (3) for all blocks of constraints. To satisfy the constraints, we calculate the error vector for block $i$ ($i \geq 2$) by equation (3) and use it in block 1 ($i = 1$) by equation (A1). Therefore, the following equation must hold for $i = 2, \ldots, n$:

$$(\mathcal{X}_1 - \mathbf{A}_1\mathbf{A}_i^{-1}\mathcal{X}_i)\boldsymbol{\psi} = \left(\mathbf{I} - \mathbf{E}_1 - \mathbf{A}_1\left(\mathbf{I} - \frac{1}{n}\mathbf{1}\right)\right)\mathbf{y}.$$

Using Proposition 2, there is $\mathbf{A}_1^{-1}(\mathbf{I} - \mathbf{E}_1) = \mathbf{I} - \frac{1}{n}\mathbf{1}$. Hence, the right hand side of the above equation is $\mathbf{I} - \mathbf{E}_1 - \mathbf{A}_1\mathbf{A}_1^{-1}(\mathbf{I} - \mathbf{E}_1) = \mathbf{0}$. Using this result, we reframe the block of constraints (A1) as $(\mathcal{X}_1 - \mathbf{A}_1\mathbf{A}_i^{-1}\mathcal{X}_i)\boldsymbol{\psi} = \mathbf{0}$. Note that the left hand side of this equation shows the violation, and the penalty is defined as sum of the quadratic violations.∎

**Theorem 3.** The following properties hold:

a) Problem (4) has an optimal solution for any given $M \geq 0$,
b) Let $\boldsymbol{\psi}^*$ and $\boldsymbol{\psi}^*(M)$ be the optimal solutions to problems (2) and (4), respectively. Then $\boldsymbol{\psi}^*(M) \to \boldsymbol{\psi}^*$ as $M \to \infty$.

**Proof.**

a) We show that matrix $\mathbf{H}$ is positive semidefinite. Matrix $\mathbf{Q}$ can be written as $\mathbf{Q} = 2\mathbf{T}'\mathbf{T}$, where $\mathbf{T} = \mathbf{A}_1^{-1}\mathcal{X}_1$. Let $\mathbf{q}$ be a nonzero vector of size $nm + n^2$. Hence, $\mathbf{q}'\mathbf{T}'\mathbf{T}\mathbf{q} \geq 0$, and therefore $\mathbf{Q}$ is positive semidefinite. Similarly, matrix **V** is positive semidefinite, and hence **H** is positive semidefinite. Therefore, for any $M > 0$ problem (4) is convex and it has a global minimizer.

b) First, assume that $\boldsymbol{\psi}^*(M)$ is a feasible solution to problem (2). This means that all the constraints are satisfied and $\boldsymbol{\psi}^{*\prime}\mathbf{V}\boldsymbol{\psi}^* = 0$. In this case, the objective function of (4) is equal to the objective function of (2) minus the scalar value of $\gamma$. Therefore, the optimal solutions of (2) and (4) are the same: $\boldsymbol{\psi}^*(M) = \boldsymbol{\psi}^*$.

Now assume that $\boldsymbol{\psi}^*(M)$ is outside of the feasible region of (2). Let $\{M_k\}$, $k = 1, 2, \ldots, \infty$ be an increasing sequence of nonnegative numbers, and let us use a more simple notation for the optimal solution of (4) as $\boldsymbol{\psi}^{(k)} = \boldsymbol{\psi}^*(M_k)$. Optimality of $\boldsymbol{\psi}^{(k+1)}$ and $\boldsymbol{\psi}^{(k)}$ implies that:

$$\frac{1}{2}\boldsymbol{\psi}^{(k)\prime}(\mathbf{Q} + M_k\mathbf{V})\boldsymbol{\psi}^{(k)} + \mathbf{c}\boldsymbol{\psi}^{(k)} \leq \frac{1}{2}\boldsymbol{\psi}^{(k+1)\prime}(\mathbf{Q} + M_k\mathbf{V})\boldsymbol{\psi}^{(k+1)} + \mathbf{c}\boldsymbol{\psi}^{(k+1)},$$

$$\frac{1}{2}\boldsymbol{\psi}^{(k+1)\prime}(\mathbf{Q} + M_{k+1}\mathbf{V})\boldsymbol{\psi}^{(k+1)} + \mathbf{c}\boldsymbol{\psi}^{(k+1)} \leq \frac{1}{2}\boldsymbol{\psi}^{(k)\prime}(\mathbf{Q} + M_{k+1}\mathbf{V})\boldsymbol{\psi}^{(k)} + \mathbf{c}\boldsymbol{\psi}^{(k)}.$$

By adding these two inequalities and rearranging the terms, we have

$$(M_{k+1} - M_k)\left(\boldsymbol{\psi}^{(k+1)\prime}\mathbf{V}\boldsymbol{\psi}^{(k+1)} - \boldsymbol{\psi}^{(k)\prime}\mathbf{V}\boldsymbol{\psi}^{(k)}\right) \leq 0,$$

As $M_{k+1} - M_k \geq 0$, there is $0 \leq \boldsymbol{\psi}^{(k+1)\prime}\mathbf{V}\boldsymbol{\psi}^{(k+1)} \leq \boldsymbol{\psi}^{(k)\prime}\mathbf{V}\boldsymbol{\psi}^{(k)}$. Hence, $\boldsymbol{\psi}^{(k)\prime}\mathbf{V}\boldsymbol{\psi}^{(k)} \to 0$ as $k \to \infty$, which means that the value of penalty term at the optimal solution to (4) decreases if $M \to \infty$. Therefore for a large $M$, $\boldsymbol{\psi}^*(M)$ is in a tight neighborhood of the feasible region of (2), and it gets closer to the feasible region if the value of $M$ increases. As a result, $\boldsymbol{\psi}^*(M) \to \boldsymbol{\psi}^*$ as $M \to \infty$. ∎

**Some discussions about big *M***

The penalty term in problem (4) is zero if the solution is feasible to problem (2). Therefore at any step of the solving process, if a feasible solution is obtained then the penalty is zero and the magnitude of $M$ is not important. However, there are some concerns for the value of $M$ in practice, which are related to the numerical precisions of the solver and the computer. Solvers usually use two tolerance thresholds for the numerical computations of a QP: optimality tolerance (OT), and feasibility tolerance (FT). The precision (PT) of the computer is also important since it causes rounding biases.

A rule of thumb for the value of $M$ is explained here. Let $\boldsymbol{\psi}$ be a solution to (4), and let $f$ be an upper limit for $\frac{1}{2}\boldsymbol{\psi}^\prime\mathbf{Q}\boldsymbol{\psi} + \mathbf{c}\boldsymbol{\psi}$ and $g$ be the value of $\frac{1}{2}M^2\boldsymbol{\psi}^\prime\mathbf{V}\boldsymbol{\psi}$. If there exists some violation then there should be $g > |f|$. Let assume there is a small violation ($\delta > FT$) in every constraint of (2). Thus $g$ is approximated by $n^2M^2\delta/2$, and there is $M > \sqrt{\frac{2|f|}{n^2\delta}}$ which approximates the lower bound of $M$. The value of $f$ can be approximated from the objective value of a linear regression problem. For example, assume $n = 100$, $FT = 10^{-8}$, $\delta = 10^{-5}$, and an approximation for $f$ is $-10^3$ (note that optimal value of $f$ is negative). Using this role of thumb, there is $M \cong 140$.

In case an extremely big $M$ is used, at the optimal solution to (4) there is $\boldsymbol{\psi}^{*\prime}\mathbf{V}\boldsymbol{\psi}^* \cong 0$ but the value of $\frac{1}{2}\boldsymbol{\psi}^{*\prime}M^2\mathbf{V}\boldsymbol{\psi}^*$ may be still larger than $\frac{1}{2}\boldsymbol{\psi}^{*\prime}\mathbf{Q}\boldsymbol{\psi}^* + \mathbf{c}\boldsymbol{\psi}^*$. If this case happens, the current solution to (4) is thus hold the concavity condition. Therefore the estimated function is piecewise linear and concave, which may be equivalent to a linear regression line, the average line (average of $y$ values), or a piecewise linear function that is not necessarily the best fit.

**Dual of penalized monotonic CLS**

Let $\boldsymbol{\mu}$ be the Lagrange multiplier of nonnegativity constraint in problem (4). We build the Lagrange function $L(\boldsymbol{\psi}, \boldsymbol{\mu}) = \frac{1}{2}\boldsymbol{\psi}'\mathbf{H}\boldsymbol{\psi} - (\boldsymbol{\mu}' - \mathbf{c})\boldsymbol{\psi}$ and minimize it in $\boldsymbol{\psi}$. This is an unconstrained optimization and the function is convex and differentiable, hence the minimum is given by $\nabla_{\boldsymbol{\psi}} L = 0$. Therefore, $\mathbf{H}\boldsymbol{\psi} = (\boldsymbol{\mu} - \mathbf{c}')$ and $\boldsymbol{\psi} = \mathbf{H}^{-1}(\boldsymbol{\mu} - \mathbf{c}')$. By substituting $\boldsymbol{\psi}$ into the Lagrange function, we get the following dual function:

$$L(\boldsymbol{\mu}) = -\frac{1}{2}(\boldsymbol{\mu} - \mathbf{c}')'\mathbf{H}^{-1}(\boldsymbol{\mu} - \mathbf{c}'),$$

and the dual problem is obtained by maximizing $L$ subject to nonnegative $\boldsymbol{\mu}$. We perform the following two steps to obtain dual of penalized monotonic CLS as presented in (6):

i. Define $\mathbf{z} = \mathbf{H}^{-1}(\boldsymbol{\mu} - \mathbf{c}')$, and obtain $\boldsymbol{\mu} - \mathbf{c}' = \mathbf{Hz}$. Hence, the dual problem is to maximize $-\frac{1}{2}\mathbf{z}'\mathbf{Hz}$ subject to $\mathbf{Hz} + \mathbf{c}' \geq \mathbf{0}$,

ii. Use $\mathbf{H} = \mathbf{F}'\mathbf{F}$ and define $\mathbf{w} = \mathbf{Fz}$. Hence, the dual problem is to maximize $-\frac{1}{2}\mathbf{w}'\mathbf{w}$ subject to $\mathbf{F}'\mathbf{w} + \mathbf{c}' \geq 0$.

To obtain the dual of penalized CLS (8), consider that the Lagrange multipliers of the first $mn$ elements of $\boldsymbol{\mu}$ are free of sign.∎

## Appendix 3.

In this appendix we present two functions: `Dual_PMCLS(x,y)` and `PMCLS(x,y)`, and one auxiliary function to generate matrix $\mathcal{X}_i$ (function `make_X`). To call the functions use the command as:

    Matlab: [SSR,eps]=Dual_PMCLS(x,y),    R: sol<-Dual_PMCLS(x,y)

where `x` is an $n \times m$ matrix of input values, `y` is a vector of outputs, `SSR` is the estimated SSR, and `eps` is the error. The function `make_X` must be in the same folder as the main functions. In R, `sol` is a list containing SSR and eps.

## Part 1. Matlab codes

### MATLAB code for dual of penalized MCLS

```matlab
function [SSR,eps]=Dual_PMCLS(x,y)
%This function refers to dual of penalized monotonic CLS
%Please cite this paper if you use this function
n= size(x,1); m= size(x,2);
ai=sparse(eye(n));
E=@(i) sparse([zeros(n,(i-1)),ones(n,1),zeros(n,(n-i))]);
ei=@(i) sparse([zeros(n,(i-1)),ai(:,i),zeros(n,(n-i))]);
ei1=@(i) sparse([zeros((i-1),n);ai(1,:);zeros((n-i),n)]);
ainv=@(i) sparse(ai-(1/n)*ones(n)+(2/n)*E(i)-ei(i));%Inverse of matrix A(i)

X=@(i) make_X(i,m,n,x,ai,1);%We make matrix X via a separate function
Q=X(1)'*sparse((ai-(1/n)*ones(n)+(1/n)*(E(1)+E(1)')-ei(1)));
Q=Q*X(1);
C=-2*y'*(ai-(1/n)*ones(n)+(1/n)*E(1)-ei(1))*X(1);

%Building matrix F
r=cell(n,1);c=cell(n,1);v=cell(n,1);
for i=2:n
    [r{i},c{i},v{i}]=find(sparse((ai-E(1)+ei1(i)'-ei(i))*X(i)-X(1)));
    r{i}=(i-1)*n+r{i};
end;
r=cell2mat(r);
c=cell2mat(c);
v=cell2mat(v); v=round(v,8);
F=sparse(r,c,v,n^2,n^2+n*m);
H= sparse(1:n^2,1:n^2,ones(n^2,1),n^2 ,n^2 );
F=100*F;
F(1:n,1:n*m+n^2)=sparse(sqrt(2)*(ainv(1)*X(1)));

% We use MOSEK to solve the problem. The reader may choose to use quadprog
% or other solvers instead.
param = [];
param.MSK_IPAR_LOG=0;
[res]=mskqpopt(H,zeros(n^2 ,1),F',[],C', [],[],param );

psi=-res.sol.itr.y;%Optimal values of psi variables
eps =sparse(eye(n)-(1/n)*ones(n,n))*y-ainv(1)*(X(1))*psi;
SSR=eps'*eps;
end
```

## MATLAB code for penalized MCLS

```matlab
function [SSR,eps]=PMCLS(x,y)
%This function refers to penalized monotonic CLS
%Please cite this paper if you use this function
n= size(x,1); m= size(x,2);
ai=sparse(eye(n));
E=@(i) sparse([zeros(n,(i-1)),ones(n,1),zeros(n,(n-i))]);
ei=@(i) sparse([zeros(n,(i-1)),ai(:,i),zeros(n,(n-i))]);
ei1=@(i) sparse([zeros((i-1),n);ai(1,:);zeros((n-i),n)]);
ainv=@(i) sparse(ai-(1/n)*ones(n)+(2/n)*E(i)-ei(i));%Inverse of matrix A(i)

X=@(i) make_X(i,m,n,x,ai,1);%We make matrix X via a separate function
Q=X(1)'*sparse((ai-(1/n)*ones(n)+(1/n)*(E(1)+E(1)')-ei(1)));
Q=Q*X(1);
C=-2*y'*(ai-(1/n)*ones(n)+(1/n)*E(1)-ei(1))*X(1);

%Building matrix F
r=cell(n,1);c=cell(n,1);v=cell(n,1);
for i=2:n
    [r{i},c{i},v{i}]=find(sparse((ai-E(1)+ei1(i)'-ei(i))*X(i)-X(1)));
    r{i}=(i-1)*n+r{i};
end;
r=cell2mat(r);
c=cell2mat(c);
v=cell2mat(v); v=round(v,8);
F=sparse(r,c,v,n^2,n^2+n*m);

V=F'*F;
H=2*Q+V*10000 ;H=round(H,8);

% We use MOSEK to solve the problem. The reader may choose to use quadprog
% or other solvers instead.
param = [];
param.MSK_IPAR_LOG=0;
[res]=mskqpopt(H,C',zeros(1,size(C,2)),[],[],zeros(size(C,2),1),[] ,param);

psi=res.sol.itr.xx;%Optimal values of psi variables
eps =sparse(eye(n)-(1/n)*ones(n,n))*y-ainv(1)*(X(1))*psi;
SSR=eps'*eps;
end
```

## MATLAB code for making matrix $\mathcal{X}$

```matlab
function X=make_X(i,m,n,x,ai,slacks)
d=@(i,p) sparse(diag(x(:,p))-x(i,p)*ai);
r=cell(m);c=cell(m);v=cell(m);
for p=1:m
    [r{p},c{p},v{p}]=find(d(i,p));c{p}=c{p}+n*(p-1);
end;
r=cell2mat(r);c=cell2mat(c);v=cell2mat(v);
X=sparse(r,c,v,n,n*m);
if slacks==1
    r=1:n;
    c=m*n+n*(i-1)+1:m*n+n*(i-1)+n;
    X2=sparse(r,c,ones(n,1),n,m*n+n^2);
    X2(:,1:m*n)=X;
    X=X2;
end;
end
```

## Part 2. R codes

### R code for dual of penalized MCLS

```r
Dual_PMCLS = function(m,n,x,y){
  # This function refers to dual of penalized monotonic CLS
  # Please cite this paper if you use this function

  require(slam)
  require(quadprog)
  source("make_X.R") #the code for this script is available in the paper
  ai <- diag(n)
  E <-function(i) return(cbind(matrix(0,n,i-1),matrix(1,n,1),matrix(0,n,n-i)))
  ei <- function(i) return(cbind(matrix(0, n, i-1),ai[, i],matrix(0, n, n-i)))
  ei1 <- function(i) return(rbind(matrix(0, i-1, n),ai[1, ],matrix(0, n-i, n)))
  ainv <- function(i) return(ai-matrix(1,n,n)/n +2*E(i)/n-ei(i))#Inverse of A(i)
  X <- function(i) make_X(i,m,n,x,ai,1)
  Q <- t(X(1)) %*% (ai-(1/n)*matrix(1, n, n)+(1/n)*(E(1)+t(E(1))-ei(1)))
  Q <- Q %*% X(1)
  C <- -2*t(y) %*% (ai-(1/n)*matrix(1, n, n)+(1/n)*E(1)-ei(1))%*%X(1)

  #Building matrix F
  r <- list(); c <- list(); v <- list()
  for (i in 2:n){
    temp <- (ai-E(1)+t(ei1(i))-ei(i))%*%X(i)-X(1)
    rc <- which(temp!=0,arr.ind = T)
    r[[i]] <- rc[,1]; c[[i]] <- rc[,2]; v[[i]] <- temp[rc];
    r[[i]] <- (i-1)*n + r[[i]];
  }
  r <- unlist(r); c <- unlist(c); v <- unlist(v);v <- round(v,8);
  F <- matrix(simple_triplet_matrix(r,c,v,n^2,n^2+n*m),n^2,n^2+n*m)
  H <-matrix(simple_triplet_matrix(1:n^2,1:n^2,matrix(1,n^2,1),n^2,n^2),n^2,n^2)
  F <- 100*F   #big M = 100
  F[1:n,1:(n*m+n^2)] <- (sqrt(2)*(ainv(1) %*% X(1)))

  # Solve by using solve.QP in R. This works but a more efficient solver such as
  # CPLEX, Gurobi or Mosel is preferred
  H <- H + diag(0.000000001,dim(H)[1],dim(H)[2])
  res <- solve.QP(H, c(rep(0,n^2)), (-F), (-C), meq=0, factorized=F)

  psi <- res$Lagrangian #Optimal values of psi variables
  eps <- (ai-(1/n)*matrix(1,n,n))%*%y-ainv(1)%*%(X(1))%*%psi;
  SSR=t(eps) %*% eps
  return(list(SSR, eps))
}
```

### R code for penalized MCLS

```r
PMCLS=function(x,y){
  # This function refers to penalized monotonic CLS
  # Please cite this paper if you use this function

  require(slam)
  require(quadprog)
  source("make_X.R") #the code for this script is available in the paper
  n <- dim(x)[1]; m <- dim(x)[2];
  ai <- diag(n)
  E <-function(i) return(cbind(matrix(0,n,i-1),matrix(1,n,1),matrix(0,n,n-i)))
  ei <- function(i) return(cbind(matrix(0, n, i-1),ai[, i],matrix(0, n, n-i)))
  ei1 <- function(i) return(rbind(matrix(0, i-1, n),ai[1, ],matrix(0, n-i, n)))
  ainv <- function(i) return(ai-matrix(1,n,n)/n +2*E(i)/n-ei(i))#Inverse of A(i)
  X <- function(i) return(make_X(i,m,n,x,ai,1))
```

```
  Q <- t(X(1)) %*% (ai-(1/n)*matrix(1, n, n)+(1/n)*(E(1)+t(E(1))-ei(1)))
  Q <- Q %*% X(1)
  C <- -2*t(y) %*% (ai-(1/n)*matrix(1, n, n)+(1/n)*E(1)-ei(1))%*%X(1)

  #Building matrix F
  r <- list(); c <- list(); v <- list()
  for (i in 2:n){
    temp <- (ai-E(1)+t(ei1(i))-ei(i))%*%X(i)-X(1)
    rc <- which(temp!=0,arr.ind = T)
    r[[i]] <- rc[,1]; c[[i]] <- rc[,2]; v[[i]] <- temp[rc];
    r[[i]] <- (i-1)*n + r[[i]];
  }

  r <- unlist(r); c <- unlist(c); v <- unlist(v);v <- round(v,8);
  F <- matrix(simple_triplet_matrix(r,c,v,n^2,n^2+n*m),n^2,n^2+n*m)
  V <- t(F)%*%F;
  H <- 2*Q+V*10000 ;H=round(H,8); # big M = 100

  # Solve by using solve.QP in R. This works but a more efficient solver such as
  # CPLEX, Gurobi or Mosel is preferred
  H <- H + diag(0.000001,dim(H)[1],dim(H)[2])
  res<-solve.QP(H,-t(C),diag(1,dim(H)[1]),c(rep(0,dim(H)[1])),
          meq=0,factorized=F)

  psi <- res$solution #Optimal values of psi variables
  eps <- (ai-(1/n)*matrix(1,n,n))%*%y-ainv(1)%*%(X(1))%*%psi;
  SSR=t(eps) %*% eps
  return(list(SSR, eps))
}
```

### R code for making matrix $\mathcal{X}$

```
make_X = function(i,m,n,x,ai,slacks=0){
  d <- function(i,p) return(diag(x[,p])-x[i,p] * ai)
  r <- list()
  c <- list()
  v <- list()
  for (p in 1:m){
    rc <- which(d(i,p)!=0,arr.ind = T)
    r[[p]] <- rc[,1]
    c[[p]] <- rc[,2]
    v[[p]] <- d(i,p)[rc]
  }
  r <- unlist(r)
  c <- unlist(c)
  v <- unlist(v)
  X <- matrix(simple_triplet_matrix(r,c,v,n,n*m),n,n*m)
  if (slacks==1){
    r <- 1:n
    c <- (m*n+n*(i-1)+1):(m*n+n*(i-1)+n)
    X2 <- matrix(simple_triplet_matrix(r,c,matrix(1,n,1),n,m*n+n^2),n,m*n+n^2)
    X2[,1:(m*n)] <- X
    X <- X2;
  }
  return(X)
}
```